\documentclass[12pt,preprint]{aastex}
\usepackage{graphicx}
\usepackage{natbib}

\shorttitle{STRANGE QUARK MATTER OBJECTS IN EXOPLANETS}
\shortauthors{Huang \& Yu}

\begin{document}

\title{Searching for Strange Quark Matter Objects in Exoplanets }

\author{Y. F. Huang\altaffilmark{1,2}, and Y. B. Yu\altaffilmark{1,2}}

\altaffiltext{1}{Department of Astronomy, School of Astronomy and Space Science, Nanjing University, Nanjing 210023, China; hyf@nju.edu.cn}
\altaffiltext{2}{Key Laboratory of Modern Astronomy and Astrophysics (Nanjing University), Ministry of Education, Nanjing 210023, China}

\begin{abstract}
The true ground state of hadronic matter may be strange quark matter (SQM).
Consequently, the observed pulsars may actually be strange
quark stars, but not neutron stars. However, proving or disproving
the SQM hypothesis still remains to be a difficult problem, due to
the similarity between the macroscopical characteristics of strange quark stars
and neutron stars. Here we propose a hopeful method to probe the existence of
strange quark matter. In the frame work of the SQM hypothesis, strange quark
dwarfs and even strange quark planets can also stably exist.
Noting that SQM planets will not be tidally disrupted even when they get very
close to their host stars due to their extreme compactness, we argue that we could
identify SQM planets by searching for very close-in planets among extrasolar
planetary systems.
Especially, we should keep our eyes on possible pulsar planets with orbital radius
less than $\sim 5.6 \times 10^{10}$~cm and period less than $\sim 6100$~s.
A thorough search in the currently detected $\sim 2950$ exoplanets around normal
main sequence stars has failed to identify any stable close-in objects that meet
the SQM criteria, i.e. lying in the tidal disruption region for normal matter
planets. However, the pulsar planet PSR J1719-1438B, with an orbital radius of
$\sim 6 \times 10^{10}$~cm and orbital period of $7837$~s, is encouragingly
found to be a good candidate.
\end{abstract}

\keywords{dense matter --- planet-star interactions --- stars: neutron
          --- planetary systems --- pulsars: general}


\maketitle

\section{INTRODUCTION}

Normal matter is constituted of electrons and nucleons. While there is still no
evidence showing that an electron can be further divided, each nucleon is found to be
composed of three up and down quarks. Pulsars are generally believed to be neutron
stars, which are mainly made up of neutrons that agglomerate together to form a highly
condensed state. With a typical mass of $\sim 1.4 M_\odot$ and a radius of only
$\sim 10$ km, the density of neutron stars can reach several times of nuclear
saturation density at the center. However, the physics of matter at these extremely high
densities is still quite unclear to us \citep{Weber05}. For example, hyperons, and baryon resonances
($\Sigma, \Lambda, \Xi, \Delta$), and even boson condensates ($\pi^-, K^-$) may appear;
quark ($u, d$) deconfinement may also happen. Especially, it has long been
suggested that even more exotic state such as the strange quark matter (SQM) may
exist inside \citep{Itoh70,Bodmer71,Witten84,Farhi84}.
Strange quark matter is constituted of almost equal numbers of $u, d$
and $s$ quarks, with the $s$ quark number slightly smaller due to its relatively
higher static mass. It has been conjectured that SQM may be the true ground state
of hadronic matter \citep{Itoh70,Bodmer71,Terazawa79}, since its energy per baryon could be less
than that of the most stable atomic nucleus such as $^{56}$Fe and $^{62}$Ni.

The existence of strange quark stars (shortened as ``strange stars'') was
consequently predicted based on the SQM hypothesis
(also known as the Bodmer-Witten hypothesis) \citep{Witten84,Farhi84}. Strange stars
could simply be bare SQM objects, or bulk SQM cores enveloped by
thin nuclear crusts \citep{Glendenning95}. The possible existence of nuclear crusts makes strange stars
very much similar to normal neutron stars for a distant observer \citep{Alcock86}, which means
it is very difficult for us to distinguish these two kinds of intrinsically distinct
stars. An interesting suggestion is that strange stars can spin at extremely short period
(less than $\sim$ 1~ms) \citep{Frieman89,Glendenning89,Friedman89,Kristian89,Madsen98,Bhattacharyya16} due
to the large shear and bulk viscosity of SQM \citep{Wang85,Sawyer89}, while the
minimum spin period ($P_{\rm spin}$) of normal neutron stars can hardly reach the submillisecond
range \citep{Frieman89,Glendenning89}.
It is thus suggested that $P_{\rm spin} < 1$~ms can be used as a criteria
to identify a strange star \citep{Kristian89}.
However, not all strange stars should necessarily
spin at such an extreme speed. Further more, the lifetime for a strange star to maintain
a submillisecond spin period should be very short even if it has an initial period
of $P_{\rm spin} < 1$~ms at birth, due to very strong electromagnetic emission of
the fast spinning dipolar magnetic field. On the technological aspect, it is also
difficult to detect submillisecond pulsars observationally.
In fact, according to the ATNF pulsar catalogue (web site: www.atnf.csiro.au/people/pulsar/psrcat), the record for
the smallest spin period of pulsars is still $\sim 1.40$ ms and only about 80 pulsars have
periods less than 3 ms among all the $\sim 2560$ pulsars observed so far.
All these factors make this method impractical at the moment.

It has also been noted that the mass-radius relations are different for these two kinds of stars.
According to the simplest MIT Bag model \citep{Farhi84,Krivoruchenko91},
it is $M \propto R^{3} $ for strange stars, but it is $M \propto R^{-3} $ for neutron
stars \citep{Baym71,Glendenning95,Avellar10,Drago14}. Unfortunately,
this method is severely limited by the fact that the masses and radii of these compact
stars cannot be measured accurately enough so far. The fact that strange stars
and neutron stars have almost similar radius at the typical pulsar mass of ${1.4 M_{\odot}}$ \citep{Lattimer07,Ozel16}
adds additional difficulties to the application \citep{Panei00}.
Several other methods have also been suggested, based on the different
cooling behaviors \citep{Pizzochero91,Page92,Lattimer94} or the
gravitational wave
emissions \citep{Madsen98,Jaranowski98,Lindblom00,Andersson02,Jones02,Bauswein10,Moraes14,Mannarelli15,Geng15}.
But either because the difference
between strange stars and neutron stars is subtle and inconclusive, or because
the practice is extremely difficult currently, we still do not have
a satisfactory method to discriminate them after more than 40 years
of extensive investigations \citep{Cheng98}.

It is interesting to note that small chunks of SQM with baryon number lower than
$10^7$ can stably exist according to the SQM hypothesis. Consequently, there is
effectively no limitation on the minimum mass of strange stars. It means the SQM
version of white dwarfs, i.e. strange dwarfs, can exist, and even strange planets
may be present in the Universe \citep{Glendenning95}.
Noting that strange planets can spiral very close to their
host strange stars without being tidally disrupted owing to their extreme
compactness, Geng et al (2015) suggested that these merger systems would serve
as new sources of gravitational wave bursts and could be used as an effective probe
for SQM. This is a very hopeful new method. The only concern is that it would take
an extremely long time for a strange planet to have a chance to merge with its host.
According to Geng et al.'s estimation, the event rate detected by even the next
generation gravitational wave experiment such as the Einstein Telescope would
not exceed a few per year \citep{Geng15}. Thus the goal is still far from being reachable
in the near future.

In this study, we suggest that we could probe the existence of SQM by searching for
close-in planets among extrasolar planetary systems. This method can significantly
increase the opportunity for success if the SQM hypothesis is correct.

\section{EXTREMELY SMALL TIDAL DISRUPTION RADIUS}

Strange planets are SQM objects of planetary masses.
They can be used to test the SQM hypothesis.
The basic idea relies on the gigantic difference between the tidal
disruption radius for an SQM planet and that for a normal matter one.

When a planet orbits around its host star, different gravitational force
(from the host star) will be exerted on different parts of the planet due
to their slight difference in distance with respect to the host. This is
the so called tidal effect. The tidal force tends to tear the planet apart,
but it can be resisted by the self-gravity of the planet when the two
objects is still far away. When the two objects approach each other,
the tidal effect will become stronger \citep{Gu03}.
There exists a critical distance, i.e. the so
called tidal disruption radius ($r_{\rm td}$), at which the tidal
force is exactly balanced by the self-gravity of the planet \citep{Hills75}.
If the distance is smaller than the tidal disruption radius ($r_{\rm td}$),
the tidal force will dominate and the planet will be completely broken up.
An analytical expression for
$r_{\rm td}$ has been derived as $r_{\rm td} \approx (6M/\pi\rho)^{1/3}$ ,
where $M$ is the mass of the central host star and $\rho$ is the
density of the planet \citep{Hills75}.

SQM planets are extremely compact and their densities are typically
$\sim 4 \times 10^{14}$ g/cm$^3$. As a result, the tidal disruption
radius for strange planets can be scaled as,
\begin{equation}
r_{\rm td}({\rm SQM})
\approx 1.5 \times 10^{6} \left(\frac{M}{1.4~M_{\odot}}\right)^{1/3} \left(\frac{\rho}{4 \times 10^{14}~\rm{g/cm}^{3}}\right)^{-1/3} \rm{cm}.
\end{equation}
We see that the tidal disruption radius for strange planets is as
small as $r_{\rm td}(\rm SQM) \sim  1.5 \times 10^6$ cm. Thus a
strange planet will retain its integrity even when it almost comes to the
surface of the central host strange star.

On the contrary, the tidal disruption radius for normal matter planet is
usually much larger. For example, for typical planets with a density of
8~g/cm$^3$, the tidal disruption radius is $\sim 8.7 \times 10^{10}$ cm.
It means typically a normal matter planet will be
disrupted at a distance of $\sim 10^{11}$ cm, and
we will in no ways be able to see a normal planet orbiting around
its host at a distance much less than this value. Even when
we take the planet density as high as 30 g/cm$^3$, the tidal disruption
radius will still be as large as $\sim 5.6 \times 10^{10}$ cm.

The analyses above remind us that we could test the SQM hypothesis
through exoplanet observations: if we detected a close-in exoplanet that lies
in the tidal disruption region for normal matter (i.e., with the orbital radius
significantly less than $\sim 5.6 \times 10^{10}$), it must be a strange
planet.

Note that when a solid asteroid (of mass $m$, radius $r$ and density $\rho$)
gets elongated in the radial direction in the centripetal gravitational
field of its host (mass $M$), the elongation stress inside
the object can also help to resist the tidal force. It will
lead to a reduced tidal disruption radius. To consider this effect,
we can approximate the elongated asteroid by a right circular cylinder
of length $2 r$. The elongation stress will be maximal at the asteroid
center, which is \citep{Colgate81},
\begin{equation}
s_{\rm c}  =  \int_{0}^{r}  \frac{2 G M}{d^3} \rho l dl = \frac{G M \rho r^2}{d^3},
\end{equation}
where $d$ is the distance between
the asteroid and the host star.
Assuming that the strength of the material is $s$ and let $s_{\rm c}$ equal
$s$, we can derive the tidal disruption radius as \citep{Colgate81}
\begin{equation}
r_{\rm td}  =  (G M \rho r^2 / s)^{1/3}
\approx 2.4 \times 10^{9} m_{18}^{2/9} s_{10}^{-1/3}  \left(\frac{\rho}{8~\rm{g/cm}^{3}}\right)^{1/9}
\left(\frac{M}{1.4~M_{\odot}}\right)^{1/3}  \rm{cm},
\end{equation}
where $m_{18} = m / 10^{18}$~g and $s_{10} = s / 10^{10}$~dyn/cm$^2$.
This equation is applicable if the elongation stress dominates over the self-gravity.
However, for an Fe-Ni planet of the Earth mass of $M_{\oplus} = 6.0 \times 10^{27}$ g,
density $\rho = 8~\rm{g/cm}^{3}$,
radius $ r = 5.6 \times 10^8 $ cm, and strength $s = 10^{10}$ dyn/cm$^2$, we find
$r_{\rm td} = 3.6 \times 10^{11} $~cm for $M = 1.4 M_{\odot}$ from the above equation.
It is significantly larger than that derived from Equation~(1).
If the density of the planet is higher (which is of more interest in our study),
the tidal disruption radius will be even larger.
Thus for relatively larger planets studied here, this effect is not significant
and can be safely omitted.

\section{EXAMINING THE OBSERVED EXOPLANETS}

Exoplanets can be detected in various ways \citep{Perryman00}.
Currently the most productive method
is through transit photometry, i.e. monitoring the periodic brightness variation
of the host star induced by the transit of a planet across the stellar disk. In this
aspect, the {\em KEPLER} mission is undoubtedly the most successful project \citep{Borucki16}.
{\em KEPLER} is a space-based optical telescope of 0.95~m aperture. It was launched in
2009 by NASA to monitor $\sim 170,000$ stars over a period of four years.
With a 105 square degree field-of-view and $\sim 10$ ppm photometry accuracy,
it successfully detected over 4600 planetary candidates and confirmed over
a thousand exoplanets. An important advantage of the transit photometry method
is that it is even possible to measure the size of the planet, so that its
density can be derived \citep{Borucki16}. In some special cases, even the
atmosphere of the planet can be probed \citep{Armstrong16}.
Another widely used method is through radial velocity measurement, which inspects
the regular radial velocity variations of the host star caused by the orbital
movement of the planet. Long-term accuracies of the host's radial motions of
several meters per second are needed, which can effectively yield all orbital
elements of the planets except the orbital inclination.
Thirdly, for pulsar planets, timing observation
is an effective method since the orbital motion of planets will affect
the arrival times of the pulsar's radio pulses. In fact, the first extrasolar
planet was detected orbiting around PSR B1257+12 just through this method \citep{Wolszczan92}.
Finally, several other not-so-commonly-used methods, such as astrometry,
gravitational microlensing, and direct imaging, have also been successfully
applied and led to the detection of a small portion of the currently known exoplanets.

Due to continuous improvements in the observational techniques above, the number of
observed exoplanets is expanding quickly in recent years \citep{Han14,Coughlin16}.
Several catalogues are available for exoplanets, such as the Exoplanet Orbit Database
(shortened as EOD hereafter)
at {\tt exoplanets.org}, the Extrasolar Planets Encyclopaedia at {\tt exoplanet.eu}, the NASA
Exoplanet Archive at {\tt exoplanetarchive.ipac.caltech.edu}, and the {\em KEPLER} exoplanet
catalogue at {\tt archive.stsci.edu/kepler}. In this study, we use the EOD database
to carry out the statistics. As to 2017 May 27, there are 5288 planets
in the catalogue, of which 2950 are confirmed planets and 2338 are candidates.
Among the confirmed planets, 322 samples are tabulated with inferred density,
and additional 2108 samples are tabulated with both mass and radius values so that their densities
can be calculated. The planet masses are available for 2937 planets, and
the orbital radii are given for 2925 samples.
With so many exoplanets in hand, we can try to search for possible SQM objects in them.

Since the planet density is a key factor that determines the tidal disruption radius,
in Figure 1 we first plot the density distribution
for all the confirmed exoplanets with densities available (2430 objects in total).
The densities of most exoplanets (about 99\% of all the samples) are less than 10 ~g/cm$^3$.
Only 4 exoplanets are listed as denser than 30~g/cm$^3$.
Note that these high-density planets (with $\rho > 30$~g/cm$^3$ ) generally have large error bars,
thus their density measurements are highly uncertain.
Figure~1 indicates that for the density of normal hadronic planets, we can take 30~g/cm$^3$ as a
reasonable upper limit.

\begin{figure}
\includegraphics[width=\linewidth]{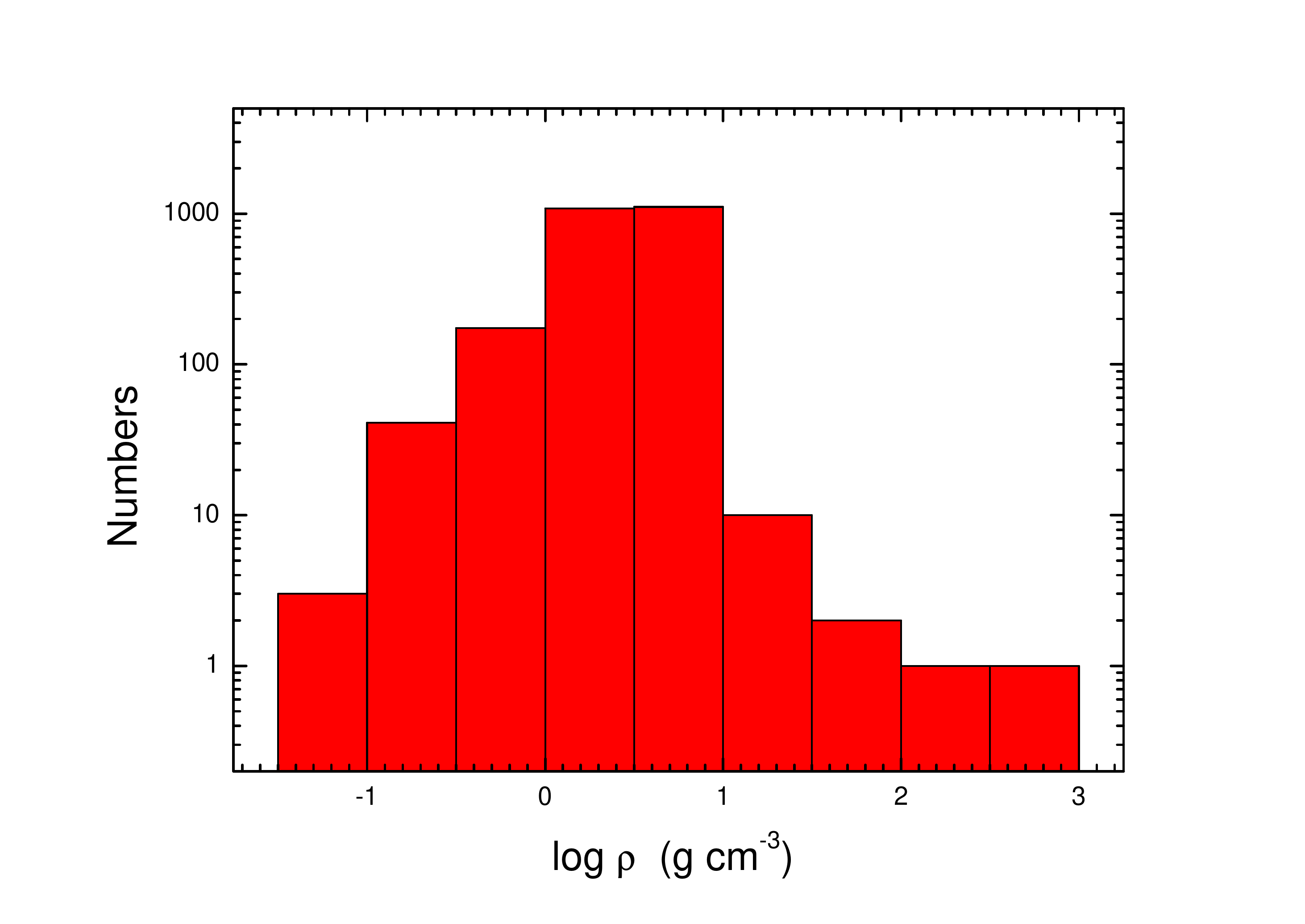}
\caption{Density distribution for all the confirmed exoplanets with a density measurement.
Only 14 planets have densities larger than 10~g/cm$^3$.
The density data are taken from the Exoplanet Orbit Database website
(http://exoplanets.org/, queried by 2017 May 27).
\label{fig1:density}}
\end{figure}

According to Equation (1), the tidal disruption radius is $r_{\rm td} \approx 5.6 \times 10^{10}$~cm
when the planet density is 30~g/cm$^3$ and the host star mass is $1.4 M_\odot$. So, a direct
strategy is to see whether there are any close-in exoplanets with the orbital radius significantly
less than the critical radius of $5.6 \times 10^{10}$~cm. In Figure~2, we plot the distribution
of orbital radius ($a$) for all the confirmed exoplanets (2925 objects in total).
Typically, the orbital radii are between 0.03~---~10 AU. For exoplanets
around normal main sequence stars, only 3 objects have
radii less than 0.01 AU. The smallest radius is 0.006 AU ($9 \times 10^{10}$~cm), but even this value
is still well above the critical tidal disruption radius of $5.6 \times 10^{10}$~cm for a
very dense object of $\rho \sim 30$~g/cm$^3$. Thus no clear clues pointing to the existence
of strange planets around normal main sequence stars are revealed from this plot.

\begin{figure}
\includegraphics[width=\linewidth]{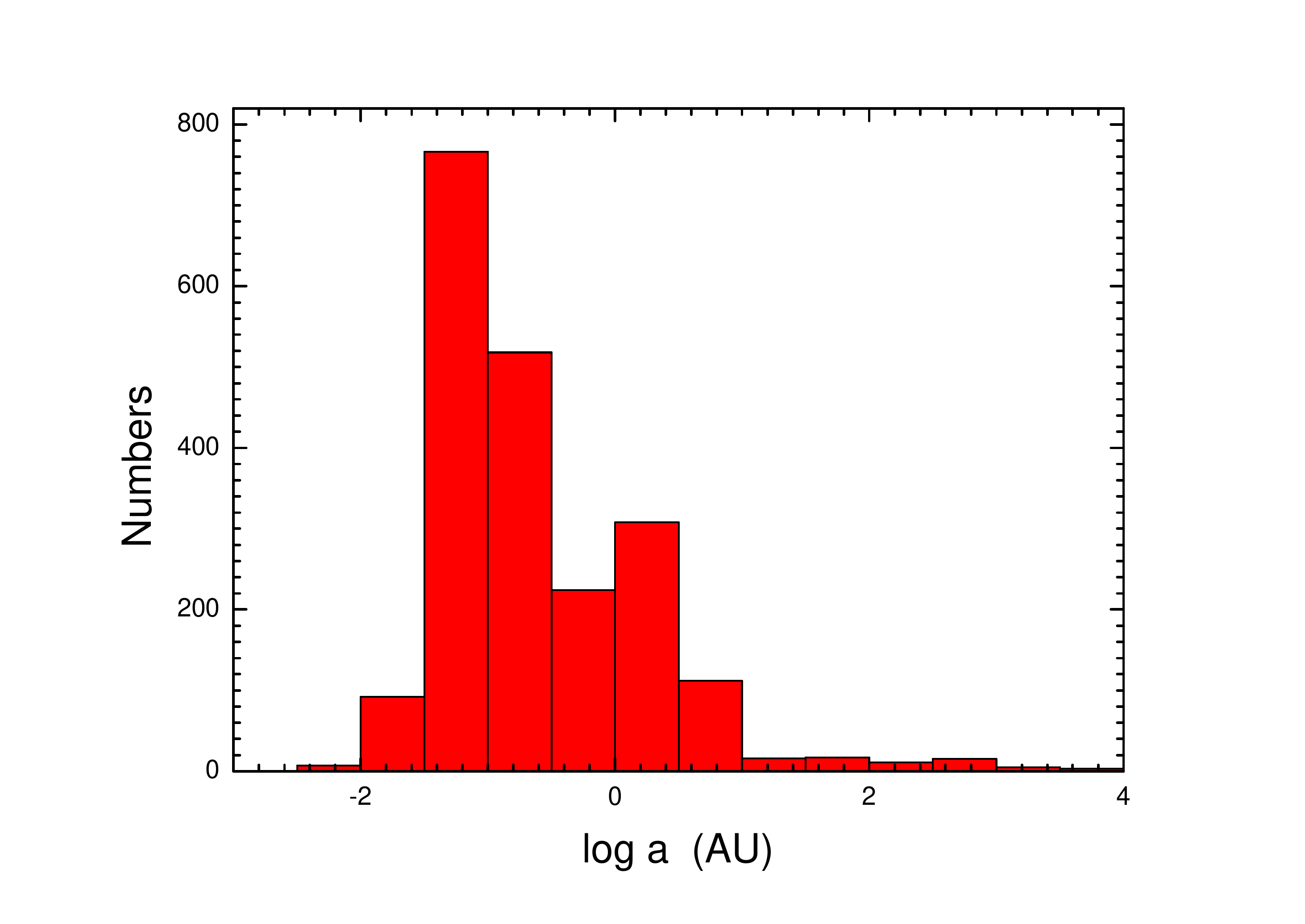}
\caption{Orbital radius distribution for all the confirmed exoplanets around
normal main sequence stars.
Most exoplanets have orbital radii larger than 0.01~AU.
The smallest orbital radius recorded currently is $0.006 $~AU ($9 \times 10^{10}$~cm).
The observational data are taken from the Exoplanet Orbit Database website
(http://exoplanets.org/, queried by 2017 May 27).
\label{fig2:distance}}
\end{figure}

Since the tidal disruption radius depends on both the planet density and the host star mass,
it is more reasonable to evaluate the closeness of planets by comparing their orbital radius
with the corresponding tidal disruption radius. We thus define the closeness of planets as $a/r_{\rm td}$.
For the planets with densities available (2430 objects, around main sequence stars), we have calculated their
tidal disruption radii ($r_{\rm td}$) and the corresponding closeness parameter.
Fig.~3a illustrates the mass distribution vs. the closeness of these planets.
It can be clearly seen that all the planets lie outside the tidal disruption region,
which proves $a > r_{\rm td}$ as a definite limitation for the survival of planets.
For the remaining 520 exoplanets without a density measurement (as listed in the
EOD database), we have assumed a typical value of 8 g/cm$^{3}$ for them and
plotted their distributions in Fig.~3b.
Again we see that no planets lie within the tidal disruption region.

From Fig.~3, it can be seen that
no clues pointing toward the existence of any SQM objects can be found in the
EOD database. This is not an unexpected result. SQM planets, if really exist, are
not likely found to be orbiting around normal main sequence stars, but
should be around compact stars (especially, strange stars). Thus we should pay
special attention to exoplanets around pulsars.
Note that for pulsar planets, the transit photometry method is not
effect and we will mainly rely on the pulsar timing method to detect them.
In this case, the densities of the planets are usually unavailable.

In fact, at least 5 planets have been detected orbiting around three
pulsars \citep{Lorimer08, Martin16}, i.e. PSR B1257+12 \citep{Wolszczan92},
PSR J1719-1438 \citep{Bailes11}, and PSR B1620-26 \citep{Backer93, Sigurdsson03}.
PSR B1257+12 has three planets, and each of the other two planets has one planetary
companion. All these planets are detected through the pulsar timing method, thus
no radius measurements are directly available for them. In Fig.~3b, we have also
plot the 5 pulsar planets, specially marked them by star symbols. Again we assumed a
typical density of 8 g/cm$^{3}$ in the plot. While four pulsar planets are safely
beyond the tidal disruption region, we do notice that one planet lies in the
disruption region (with $a/r_{\rm td} = 0.69$). It is
associated with PSR J1719-1438, a 5.7-millisecond pulsar, with an orbital radius
of $\sim 6.0 \times 10^{10}$ cm and orbital period of $\sim 2.2$ hours.
Interestingly, this problem has already been noticed by Bailes \emph{et~al.},
who argued that this companion must be denser than 23 g/cm$^{3}$ to survive the
strong tidal force of its host \citep{Bailes11}. They even went further to suggest that
the planetary companion may actually be a carbon white dwarf. However, with a mass
comparable to that of the Jupiter, it will be too rare for a white dwarf to have such
an ultralow mass. A more reasonable suggestion has been made by Horvath, who argued
that it must be an exotic quark object \citep{Horvath12}. Our current study strongly
supports Horvath's suggestion, i.e. the planet of PSR J1719-1438 is a possible SQM candidate.
It is thus very encouraging that while only 5 pulsar planets are detected, we already have one SQM
candidate among them. It hints us that close-in exoplanets would be a hopeful and
powerful tool to test the SQM hypothesis.

\begin{figure}
   \includegraphics[width=\linewidth]{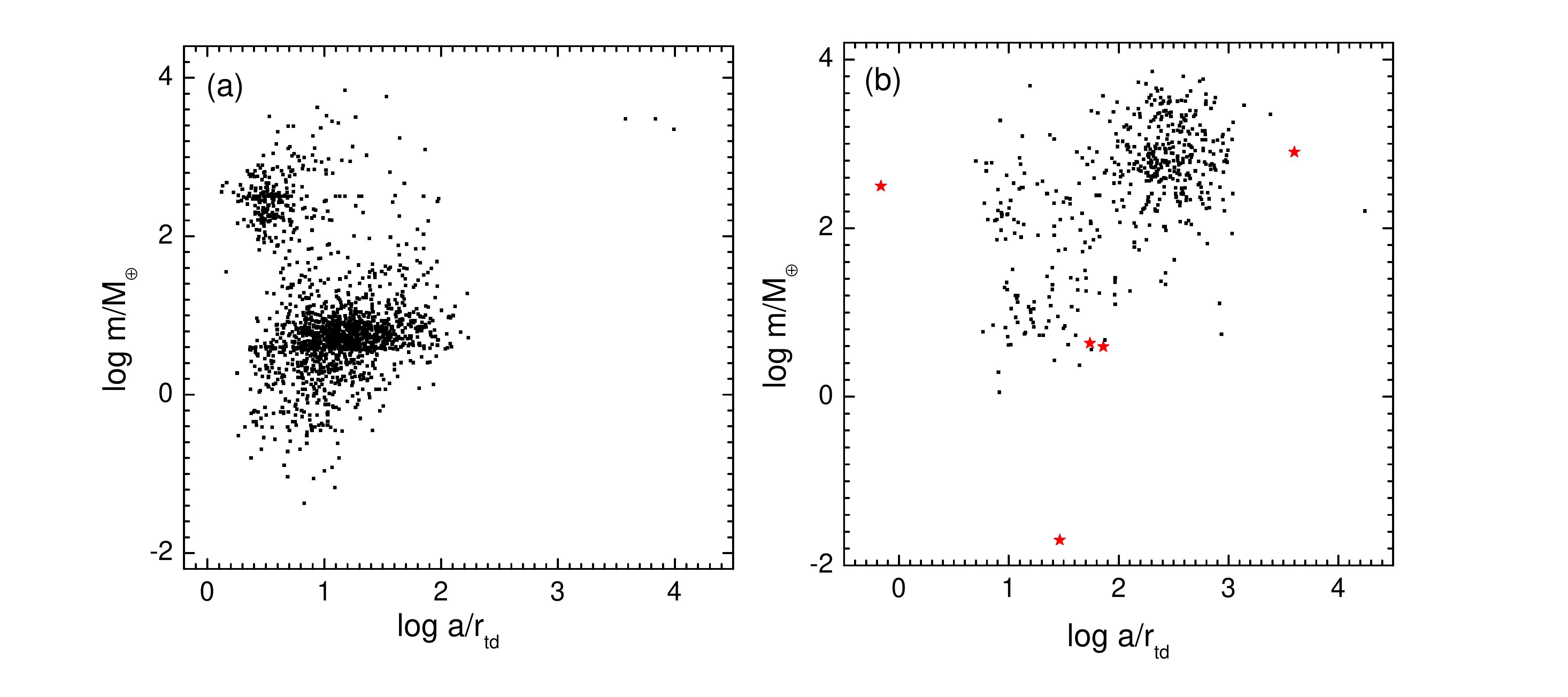}
   \caption{Closeness of all the confirmed exoplanets with respect to their hosts.
   The Y-axis shows the planet masses, and the X-axis shows their orbital radii
   in units of the corresponding tidal disruption radius.
   The exoplanet data are taken from the Exoplanet Orbit Database website at {\tt exoplanets.org}.
   \textbf{Panel (a)}: this panel is plot for the exoplanets with their densities measured.
   \textbf{Panel (b)}: for those exoplanets with densities unavailable, we assume a typical
   iron-nickel density of 8~g/cm$^{3}$ to calculate the tidal disruption radius.
   Note that the 5 star symbols stand for planets around pulsars. }
   \label{Fig3:closeness}
\end{figure}

\section{DETECTABILITY OF CLOSE-IN PULSAR PLANETS}

Searching for close-in exoplanets around pulsars should be the main direction of our future efforts.
Due to their extreme closeness, these planets will only exert a very small radial
velocity perturbation on the central compact host, which will be difficult to
be found by pulsar timing observations. Next, we give an estimate on the lower
mass limit of the planets that could be detected with current
observational techniques.

Let us consider a planet of mass $m$ orbiting around a pulsar ($M$).
In half of the orbital period, the pulsar will have a positive radial velocity
perturbation with respect to us, owing to the existence of the small companion,
while in the other half orbit, it has a negative velocity perturbation. As a
result, the topocentric time-of-arrival (TOA) of its clock-like pulses will
systematically deviate from normal rhythm regularly. The accumulated
TOA deviation can be as large as several milliseconds in each of the half orbit
and can be potentially detected through long-term timing observations.
In fact, assuming a circular orbit, the planet mass is connected with
the semi-amplitude $\Delta t$ of the
corresponding TOA variations as \citep{Wolszczan92,Wolszczan12}
\begin{equation}
m\sin{i} \approx
21.3~M_{\oplus}~\left(\frac{\Delta t}{1~{\rm ms}}\right)~\left({\frac{P_{\rm orb}}{1~{\rm day}}}\right)^{-2/3}
~\left(\frac{M}{1.4~{M_\odot}}\right)^{2/3},
\end{equation}
where $P_{\rm orb}$ is the planet's orbital period, $i$ is the orbital inclination,
and $M_{\oplus} = 6.0 \times 10^{27}$ g is the Earth mass.

The pulsar timing method essentially is also trying to measure the radial velocity
perturbation. By accumulating the TOA residuals induced by the radial velocity variation
in half of the orbit and with the microsecond precision of timing observations, it can
equivalently measure the radial velocity perturbation at an unprecedented accuracy
of $\sim 1$ cm/s. As a contrast, traditional radial velocity measurement through
optical spectroscopy can only achieve an accuracy of $\sim 1$ m/s currently.
Timing observation is thus an ideal method that could be effectively used to search for
possible close-in strange planets around pulsars.

In view of the radial velocity variation ($\Delta V$) of the host pulsar, Equation (4)
can be conveniently expressed as
\begin{equation}
m\sin{i} \approx (M a/G)^{1/2} \Delta V
\approx 0.0034 M_{\oplus} \left(\frac{M}{1.4~M_{\odot}}\right)^{1/2} \left(\frac{a}{10^{10} \rm{cm}}\right)^{1/2} \frac{\Delta V}{1 \rm{cm/s}}  ,
\label{mv}
\end{equation}
where $G$ is the gravitational constant.
Taking 30~g/cm$^3$ as a secure upper limit for the density of typical normal planets, we get
the critical tidal disruption radius as $r_{td} \approx 5.6 \times 10^{10} $~cm (Section 2).
We thus need to search for strange planets with
orbital radii smaller than this value. In fact, all the currently detected exoplanets
(except the pulsar planet PSR J1719-1438B) lie far beyond this region (Fig.~2).
From Equation (5), we see that at the
limiting radius ($a \sim 5.6 \times 10^{10} $ cm),
all planets more massive than $\sim 0.008 M_{\oplus}$ can be detected by
current pulsar timing observations.
For more close-in strange planets, even less massive
SQM planets can also be detected. Taking typical values of $i = 45^{\rm o}$, $\Delta V = 1$ cm/s,
and $M = 1.4$ M$_\odot$, we have plot the limiting mass of planets that could be detected
in Fig.~4. The figure gives us the encouraging information that close-in strange planets
need not to be very massive to be detected with our current observational techniques.

\begin{figure}
\includegraphics[width=\linewidth]{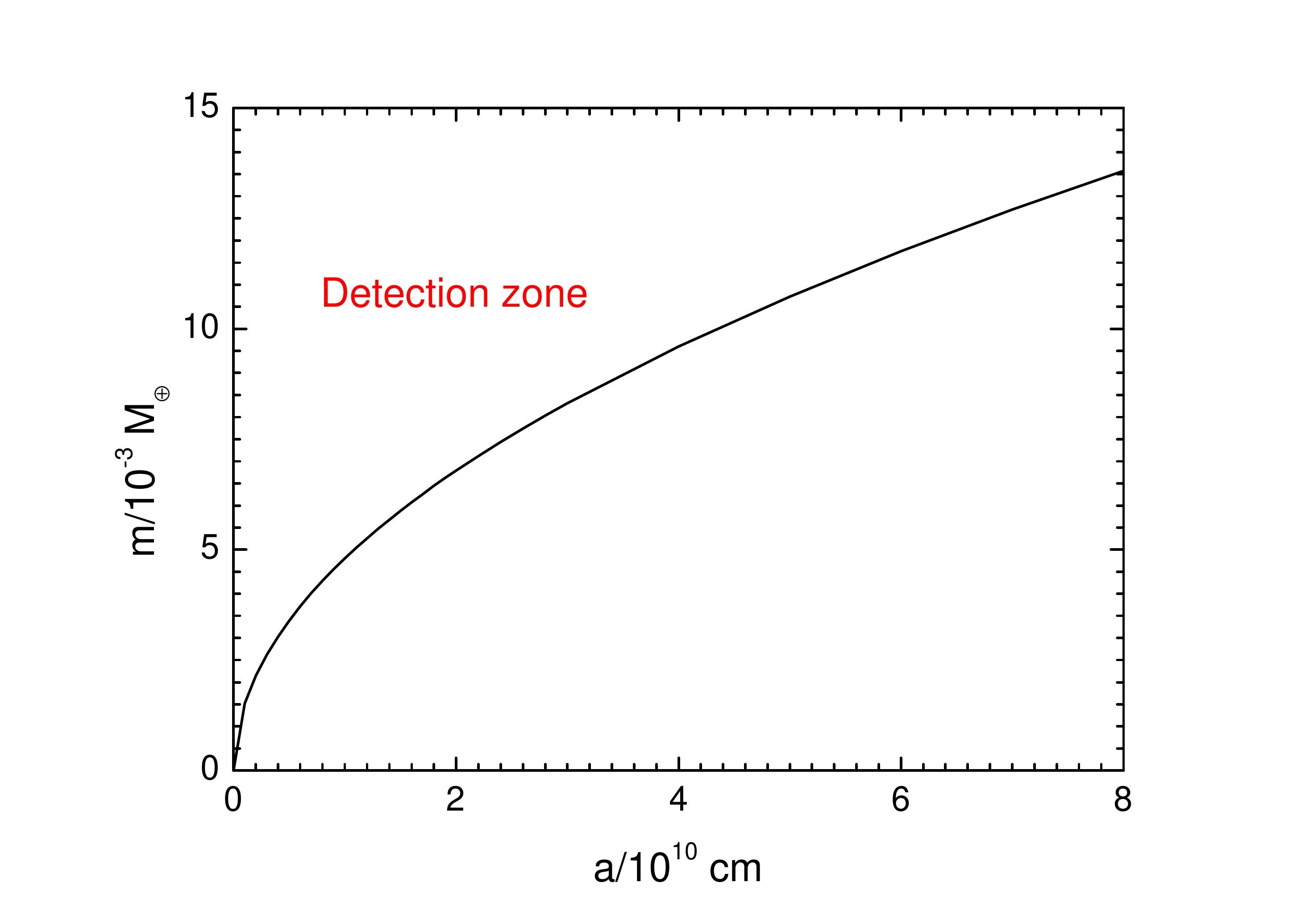}
   \caption{Minimum masses of close-in planets that could be detected by the current pulsar timing observations.
   We assume a velocity measurement accuracy of 1 cm/s for the host star.
   The orbital inclination angle is taken as $45^{\rm o}$ and the pulsar mass is taken as 1.4 $M_{\odot}$ in this plot.}
   \label{Fig4:minimass}
\end{figure}

Lying in the tidal disruption region for normal matter, these strange planets will also have very small
orbital periods. According to the Kepler's law, the radius and period of the orbit are
related by
\begin{equation}
\frac{a^3}{P_{\rm orb}^2} \approx   \frac{G M}{4 \pi^2}.
\end{equation}
At the limiting radius of $r_{td} \approx 5.6 \times 10^{10} $ cm, the period is
$P_{\rm orb} \approx 6100$~s. For more close-in orbits, the periods will be even smaller.
In Fig.~5, the relation between $P_{\rm orb}$ and $a$ is plotted for these close-in orbits.
From this figure, we see that in addition to the criterion of $a < 5.6 \times 10^{10}$ cm,
the small orbit period of $P_{\rm orb} < 6100$~s is another specific feature for SQM planets.
PSR J1719-1438B has an orbital radius of $\sim 6 \times 10^{10}$~cm and orbital period of $7837$~s.
Its orbital parameters are slightly above the SQM criteria, but it still can be regarded as
a good candidate.

\begin{figure}
\includegraphics[width=\linewidth]{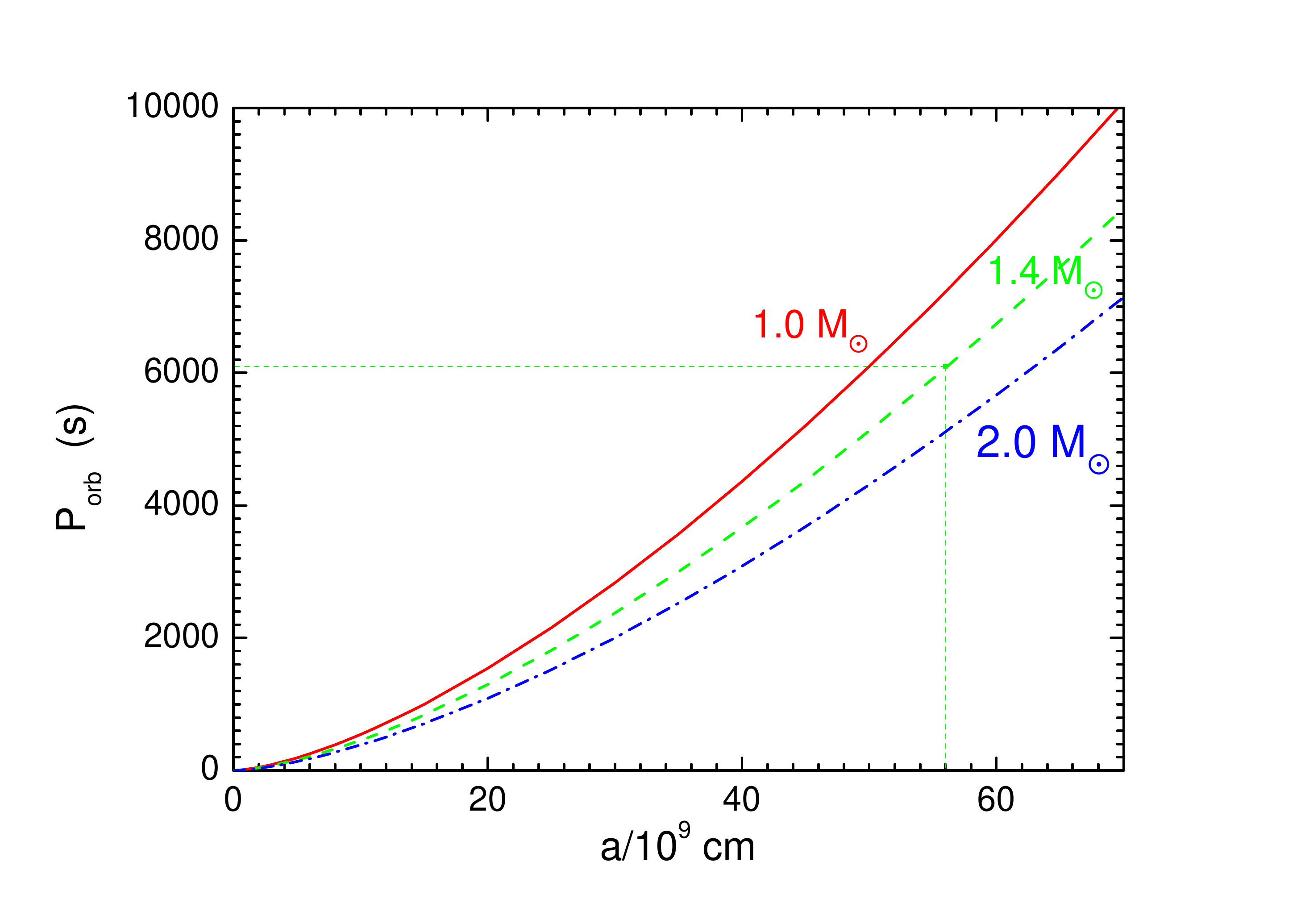}
   \caption{Orbital period vs. orbital radius for close-in planets. This figure is plotted
   according to Equation~(4). The mass of the host pulsar is taken as 1.0~$M_{\odot}$ (the solid line),
   1.4~$M_{\odot}$ (the dashed line), and 2.0~$M_{\odot}$ (the dash-dotted line), respectively.
   For the limiting orbital radius of $a = 5.6 \times 10^{10}$~cm, the corresponding orbital period is
   $\sim 6100$~s in the 1.4~$M_{\odot}$ pulsar case (see the vertical and horizontal dotted lines). }
   \label{Fig5:minip}
\end{figure}

\section{CONCLUSIONS AND DISCUSSION}

Discriminating strange stars from neutron stars observationally is an important but
challenging problem \citep{Cheng98,Xu01,Weber05,Bauswein09,Adriani15,Drago16}.
A few possible methods have previously been suggested in
the literature, but they are either inconclusive or impractical currently.
We here propose a unique method to test the SQM hypothesis: searching for close-in
exoplanets with very small orbital radius ($a < 5.6 \times 10^{10}$ cm) and very small orbital period
($P_{\rm orb} < 6100$~s). It is based on the fact that SQM planets are extremely compact and
can survive even when they are in the tidal disruption region for normal hadronic planets.
We have examined all the detected exoplanets around main sequence stars and
found no clues pointing toward the existence of SQM objects among them.
However, the pulsar planet PSR J1719-1438B, which has an orbital radius of
$\sim 6 \times 10^{10}$~cm and orbital period of $7837$~s, is found to be an interesting
candidate.

We stress that in the future, such efforts should be made mainly on exoplanets around pulsars,
since SQM planets are most likely associated with such compact stars (which themselves should
also be strange quark stars in this case). Theoretically, SQM planets can be formed
in a few ways. First, at the birth of an SQM star (either from the phase transition of
a massive neutron star, or from the merge of two neutron stars), plenty of small SQM nuggets
should be ejected. These SQM nuggets will ``contaminate'' the surrounding normal planets and
convert them into SQM planets. It means that if the Bodmer-Witten hypothesis is correct so that
neutron stars are actually strange stars, then strange planets should also be quite common.
Second, SQM clumps of planetary masses may be ejected from a strange quark star at its birth,
because the newly formed SQM host star should be hot and highly turbulent, giving birth to
high-velocity eddies \citep{Xu03, Horvath12}. These clumps may finally become planets around
the host star due to its deep gravitational potential well. Interestingly, the SQM planets
formed in this way are most likely close-in, since the ejection may not be too fierce.
Third, planetary SQM objects may be directly formed at an
early stage of our Universe, i.e. the so-called quark phase stage, when the mean density of
the Universe is extremely high \citep{Cottingham94}. Some of these SQM objects may survive
and be captured by compact stars (and even by main sequence stars) to form planetary
systems at later stages. With an unprecedented equivalent radial velocity
accuracy of $\sim 1$ cm/s, the pulsar timing method could reveal close-in planets as small
as $\sim 10^{-2} M_{\oplus}$. We appeal to radio astronomers to pay more attention on
searching for such close-in exoplanets in the future. If found, it will lead to a final
solution for the long-lasting and highly disputed fundamental problem.

\begin{acknowledgments}
The authors thank Jin-Jun Geng and Li-Jun Gou for helpful discussions.
This work was supported by the National Natural Science Foundation of China with
Grant No. 11473012, by the National Basic Research Program of China with
Grant No. 2014CB845800,
and by the Strategic Priority Research Program of the Chinese Academy of Sciences
``Multi-waveband Gravitational Wave Universe'' (Grant No. XDB23040000).
This research has made use of the Exoplanet Orbit Database
and the Exoplanet Data Explorer at {\tt exoplanets.org}.
\end{acknowledgments}

\label{lastpage}
\end{document}